\documentclass[fleqn,10pt]{wlscirep}
\usepackage[utf8]{inputenc}
\usepackage[T1]{fontenc}
\usepackage{multirow}
\usepackage{graphicx}
\title{Task-dependent fractal patterns of information processing in working memory}

\author[1,2,*]{Jeremi K. Ochab}
\author[3,1]{Marcin W\c{a}torek}
\author[4]{Anna Ceglarek}
\author[4]{Magdalena F\c{a}frowicz}
\author[4]{Koryna Lewandowska}
\author[4]{Tadeusz Marek}
\author[4]{Barbara Sikora-Wachowicz}
\author[5,1,**]{Paweł Oświ\c {e}cimka}

\affil[1]{Institute of Theoretical Physics, Jagiellonian University, 30-348 Kraków, Poland}

\affil[2]{Mark Kac Complex Systems Research Centre, Jagiellonian University, 30-348 Kraków, Poland}

\affil[3]{Faculty of Computer Science and Telecommunications, Cracow University of Technology, 31-155 Kraków, Poland}

\affil[4]{Department of Cognitive Neuroscience and Neuroergonomics, Jagiellonian University, 30-348 Kraków, Poland}

\affil[5]{Complex Systems Theory Department, Institute of Nuclear Physics, Polish Academy of Sciences, 31-342 Kraków, Poland}

\affil[*]{jeremi.ochab@uj.edu.pl}
\affil[**]{pawel.oswiecimka@ifj.edu.pl}


\begin{abstract}
We applied detrended fluctuation analysis, power spectral density, and eigenanalysis of detrended cross-correlations to investigate fMRI data representing a diurnal variation of working memory in four visual tasks: two verbal and two nonverbal. We show that the degree of fractal scaling is regionally dependent on the engagement in cognitive tasks. A particularly apparent difference was found between memorisation in verbal and nonverbal tasks. Furthermore, the detrended cross-correlations between brain areas were predominantly indicative of differences between resting state and other tasks, between memorisation and retrieval, and between verbal and nonverbal tasks. 
The fractal and spectral analyses presented in our study are consistent with previous research related to visuospatial and verbal information processing, working memory (encoding and retrieval), and executive functions, but they were found to be more sensitive than Pearson correlations and showed the potential to obtain other subtler results.
We conclude that regionally dependent cognitive task engagement can be distinguished based on the fractal characteristics of BOLD signals and their detrended cross-correlation structure.

\end{abstract}

\begin{document}

\flushbottom
\maketitle
%
%
\thispagestyle{empty}

\section*{Introduction}
\label{intro} 

Working memory is the foundation for goal-directed behaviours in which short-term storage and manipulation of sensory information are crucial for successful task execution. In understanding working memory as a core cognitive process it is essential to investigate the selection, initiation, and termination of information-processing, in other words, stages of the memory process. In our study we utilised one of the most widely used  experimental  method for memory investigation -- the Deese–Roediger–McDermott (DRM) paradigm~\cite{deese_prediction_1959,roediger_creating_1995} in which the temporal subprocesses of working memory such as encoding and retrieval are separated. This paradigm is a popular research tool  because of its methodological simplicity and controllability. Originally used to study long-term memory, it was later adapted to the working memory and functional magnetic resonance imaging (fMRI) environment~\cite{slotnick_2004, atkins_neural_2011}, allowing measurement of neural activity while a person memorises and retrieves information. In the classical DRM paradigm participants are presented with a list of semantically related words at the encoding stage but the paradigm has been successfully used with other stimuli  which allowed us to employ the modified versions of the DRM paradigm in four experimental visual tasks: two verbal and two nonverbal. 
Studies on neural correlates of working memory revealed mainly activations of the prefrontal or visual regions~\cite{pessoa_2002}. 
In recent years, functional activity in general has been intensely analysed with a variety of methods~\cite{lurie_questions_2020}.
However, blood oxygen-level-dependent (BOLD) signals have a nontrivially associated autocorrelation and cross-correlation structure~\cite{ochab_pros_2019} and remain notoriously challenging to analyse due to their very low temporal resolution.
Motivated by prior work, we applied the fractal methodology to test for regional differences in BOLD scaling properties between the tasks and experimental phases of a working memory experiment. 

Many natural systems representing diverse scientific disciplines such as physics, chemistry, biology, economics, and social sciences can be considered complex systems~\cite{kwapien_complex_2012}. These systems consist of numerous nonlinearly interacting elements and reveal extremely complex dynamical behaviour challenging to characterise by the standard analytic tools. Despite their complexity, the systems reveal some universal features, with fractality being one of the most intriguing effects among them.
It is the consequence of scale-free -- multiscale -- data/system organisation 
and, thus, the system's complexity can be quantitatively measured by fractal dimensions.
Moreover, recognising the fractal structure can manifest the system's underlying process, like self-organised criticality or cascade processes~\cite{chialvo2004,hesse2014}. A clear example is a human brain, for which fractal organisation has been identified in morphological and physiological properties~\cite{johnson2019,kiselev2003}. Moreover, in recent years this methodology, as very effective in detecting long-range temporal correlations, has also been developed in the cross-correlation case and used to quantify the nonlinear coupling between time series in economic systems\cite{oswiecimka_detrended_2014}.  

However, the proper quantification of the temporal organisation of the data is a subtle task, especially when the data are nonstationary~\cite{oswiecimka_effect_2013}. For fractals, the considered data's correlation structure obeys a power law which denotes long-range temporal or spatial correlations. Thus, the system (or signal) properties are evaluated on different scales to detect such correlations. To do this, one can use the autocorrelation function with correlations estimated at different time lags. However, determining the correlation in this way is problematic due to the noise superimposed on the data collected~\cite{kantelhardt2001,holl2015}. As a result, such data nonstationarity can lead to inaccurate estimation of the scaling exponents. Thus, one is interested in methodology, eliminating the trends without knowing their origin and shape. In this respect, two approaches are commonly used~\cite{kantelhardt2001}:
one related to the investigation of the signal in the frequency domain and the other directly in the time domain. In the former case, analysis of the signal wavelet transform reveals the correlation structure of the data at different levels of temporal resolution. One can also use Fourier transform methods to assess the scaling exponents as a complementary method. However, in this case, the nonstationarity can affect the calculations. In the latter approach, through estimating the detrended variance on different time scales, one can evaluate the Hurst exponents and quantify the level of data persistency.
  
Fractal-like or spectral methods can also help capture the information hidden in physiological signals. These methods are used successfully in the study of EEG signals~\cite{finotello_2015, marino_2019,ville_2010}, identifying the signature of scale-free dynamics.
The temporal characteristics of the fMRI signals have also been demonstrated to follow the power-law relationship expressed in the frequency domain $f$ in the form $P(f) \propto 1/f^\beta$ with $\beta>0$~\cite{zarahn1997NEmpirical} and the spatiotemporal functional correlations have been related to the dynamics of a critical system~\cite{fraiman2012FPWhat}.
Recently, also the fractal methodology was also applied to analyse fMRI data~\cite{porcaro_2020}.
Moreover, several studies showed that the fractal properties can vary between brain regions as well as during task activation~\cite{he_2011,barnes_2009, churchill_2016}.

Our study focused on the temporal organisation of data related to involvement in various cognitive tasks. To this end, we applied the fractal methodology to investigate the correlation structure of the time series related to four different experimental visual tasks: nonverbal (abstract objects requiring global or local information processing) and verbal (words or pseudowords), phases (information encoding and retrieval) and times of day of a working memory experiment. Two well-established methods of estimating the self-affinity parameter representing power-law autocorrelation signal properties, namely detrended fluctuation analysis and power spectrum density, have been applied in this context. Moreover, the nonlinear cross-correlation coefficient has been used to quantify the strength of the coupling between the time series under investigation. Our results confirm that both fractal characteristics of the time series organization and cross-correlation between brain areas are strongly associated with the brain activity related to engagement in a given task.

\begin{figure}
\centering
{
\includegraphics[scale=0.4]{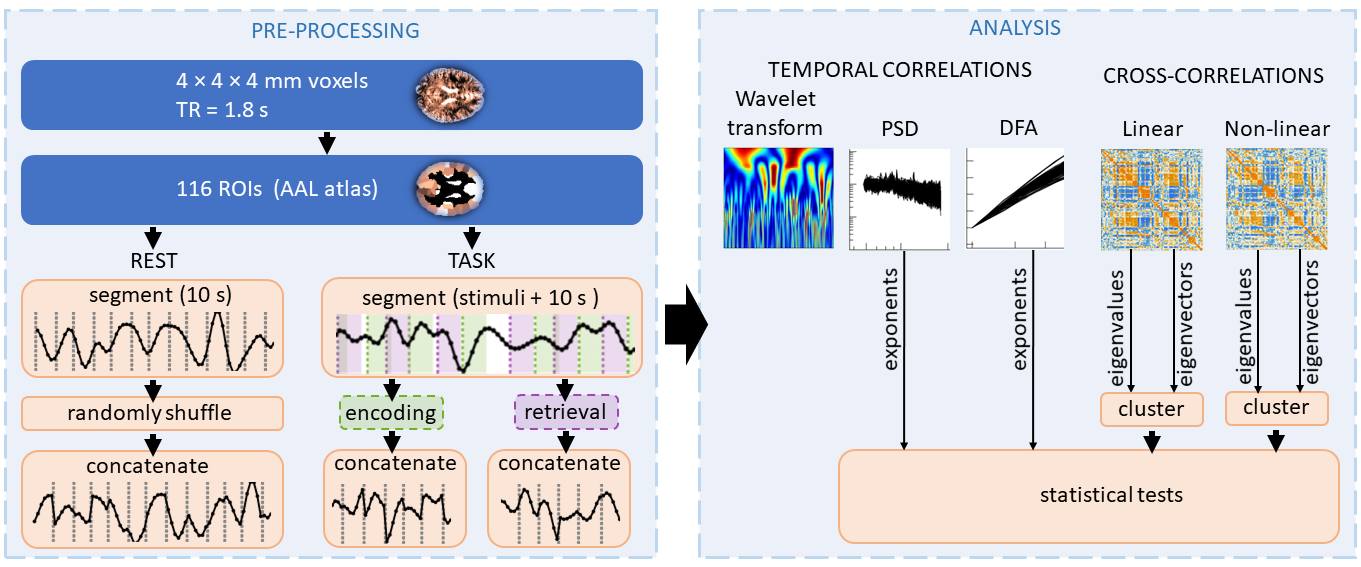}
}
\caption{Flow chart of data preprocessing and analysis performed as described in detail in Sec.~\ref{sec:Data}-\ref{sec:meth}.}
\label{fig1:flowchart}
\end{figure}

The organisation of the paper is as follows; in Section~\ref{sec:Data} we provide a summary of data specification and preprocessing.
Section~\ref{sec:meth} presents the methods of fractal, spectral, and cross-correlation analysis; a brief summary of the consecutive steps of data preprocessing and analysis is depicted in Fig.~\ref{fig1:flowchart}.  
Results of the analyses are presented in Section~\ref{sec:res}.
Finally, in Section~\ref{sec:con} the findings are discussed and conclusions presented.

\section{Results}
\label{sec:res}

\subsection{Analysis of the fractal organisation of the time series}
\label{sec:res_H}

\begin{figure}
\centering
\resizebox{.67\textwidth}{!}{
\includegraphics{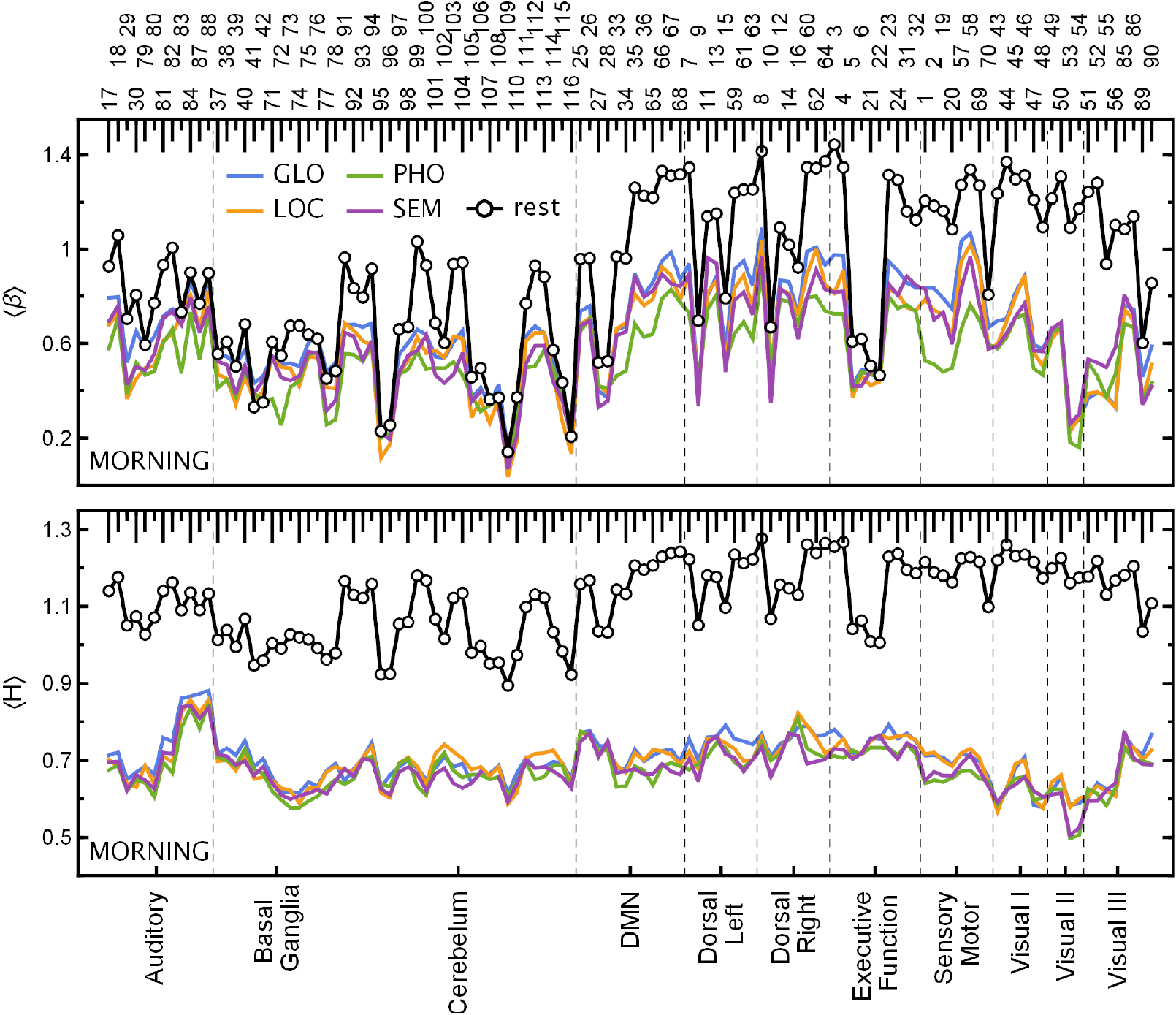}
}
\caption{Plots of the spectral, $\beta$, and Hurst, $H$, scaling exponents estimated for the entire time series. The exponents were calculated for each ROI and ordered on the plot according to the AAL atlas (top labels) and resting-state networks (bottom labels). For evening session results, see Supplementary Fig.~A.2. 
Abbreviations: visual-verbal tasks (semantic, SEM, and phonological, PHO), visual-nonverbal tasks (local information processing, LOC, and global information processing, GLO), resting-state (rest).
}
\label{fig4:Hurst_entire_data}
\end{figure}

At the beginning, we estimated the scaling properties of the time series extracted from the specific brain regions (ROIs; regions of interest) in four experimental tasks (visual-verbal: semantic, SEM, and phonological, PHO, and visual-nonverbal: local information processing, LOC, and global information processing, GLO) by applying both the DFA and PSD presented in Section~\ref{sec:meth}. Examples of the analysed data are depicted in Section~\ref{sec:Data}. For each session, we estimated the fluctuation function $F(s)$ and the power spectral density $S(f)$ of the time series in each ROI in a given task.
The corresponding quantities were averaged over all sessions, yielding the averages $\langle F(s)\rangle$ and $\langle S(f) \rangle$ for each ROI and task separately.
Based on the power-law dependence they clearly follow, as shown in detail in Section~\ref{sec:meth_H},
the Hurst and $\beta$ exponents were estimated for both tasks and the resting state and depicted in Fig.~\ref{fig4:Hurst_entire_data}.
The difference in these exponents between the resting state and the tasks is clear.
The extreme values of the scaling exponents were obtained for the resting state with $\beta \in [0.1, 1.4]$ and $H\in [0.9, 1.2]$.
In most cases, the theoretical relation between $\beta$ and $H$ (\ref{eq:relation_H_beta}) was fulfilled.
Still, we note that in some ROIs the $\beta$ exponent dropped and assumed low values, which was not observed for the Hurst exponent.
This is due to the nonstationarity of the signal, which is removed in the DFA-based methods but can strongly influence the $\beta$ values.

The estimated exponents indicate strongly correlated signals similar to the $1/f$ process, which is a ubiquitous phenomenon in physical and biological processes.
However, the signals corresponding to task performance were much less autocorrelated and characterised by exponents $\beta$ and $H$ within the range between $[0.2,0.8]$ and $[0.49,0.9]$, respectively.
Moreover, the variation of the exponents for different ROIs was similar for all kinds of task.
In figures such as Fig.~\ref{fig4:Hurst_entire_data}, we group ROIs belonging to the same resting-state networks (RSN) for greater interpretability.
The Hurst exponent assumed the highest values -- suggesting a strong persistence of the signals -- for the ROIs belonging to the Auditory RSN.
On the other hand, the smallest $H$ values characterise Visual II RSN.

To investigate possible differences in the fractal signal properties recorded in subsequent experimental phases, we extracted the time series separately from the memory encoding and retrieval phases, as described in Fig.~\ref{fig1:flowchart} (see also examples in Section~\ref{sec:Data_Preprocessing}).
The PSD and DFA characteristics were then estimated and compared with respect to the time of day when the experiment was performed (morning, evening), the tasks and the phases of each task.

\begin{figure}
\centering
\includegraphics[width = 0.8 \textwidth]{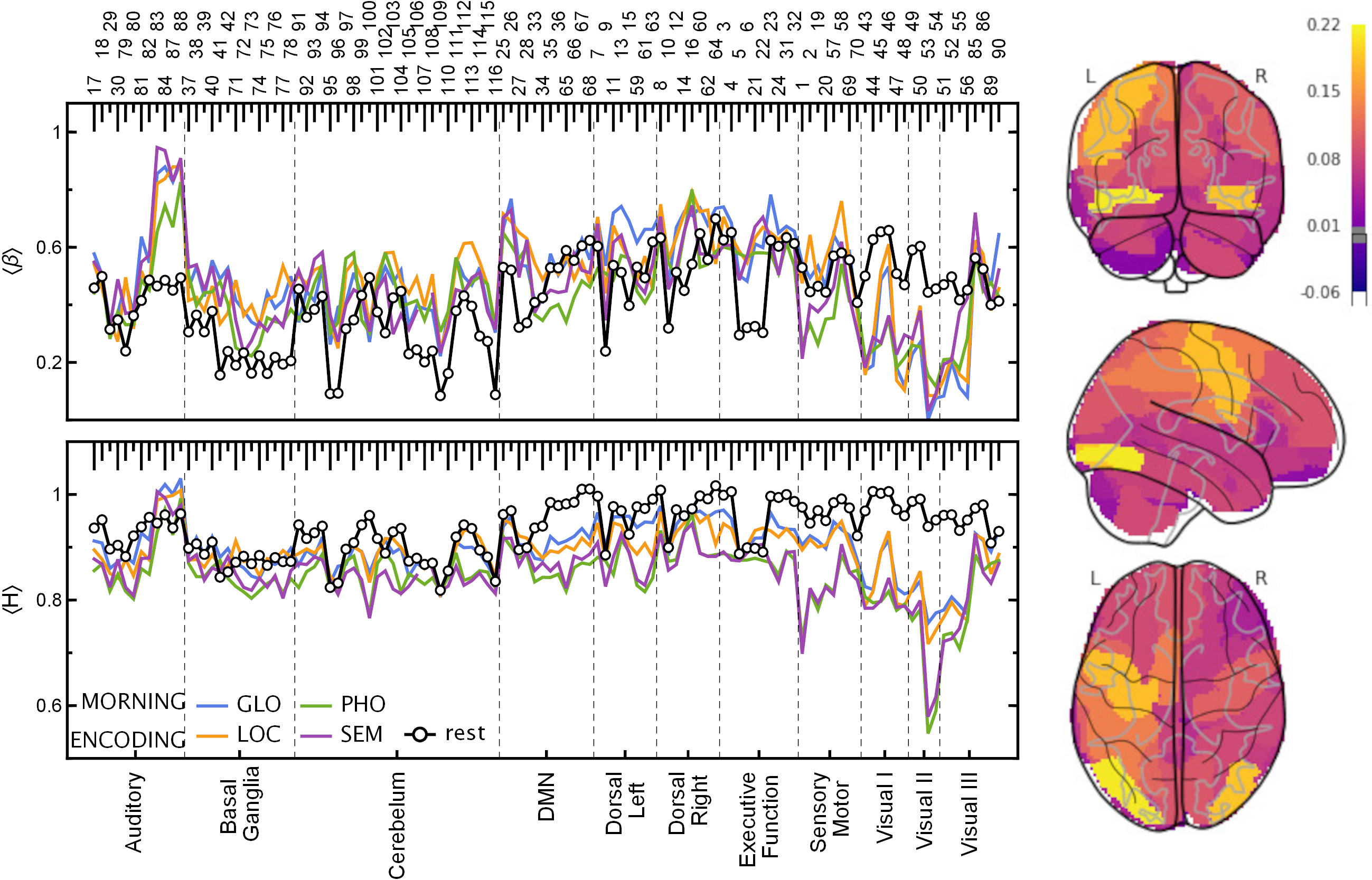}
\caption{(Left panel) Plots of the spectral, $\beta$, and Hurst, $H$, scaling exponents estimated for time series in the encoding phase. The exponents were calculated for each ROI and ordered on the plot according to the AAL atlas (top labels) and resting-state networks (bottom labels). (Right panel) Glass brain plot of $\langle H \rangle$ differences between the GLO and PHO tasks rendered with \textit{nilearn}~\cite{nilearn,nilearn_soft,scikit-learn}. For evening session results, see Supplementary Fig.~A.3. 
Abbreviations: visual-verbal tasks (semantic, SEM, and phonological, PHO), visual-nonverbal tasks (local information processing, LOC, and global information processing, GLO), resting-state (rest).
}
\label{fig5:Hurst_Enc}
\end{figure}
\begin{figure}
\centering
\includegraphics[width = 0.8 \textwidth]{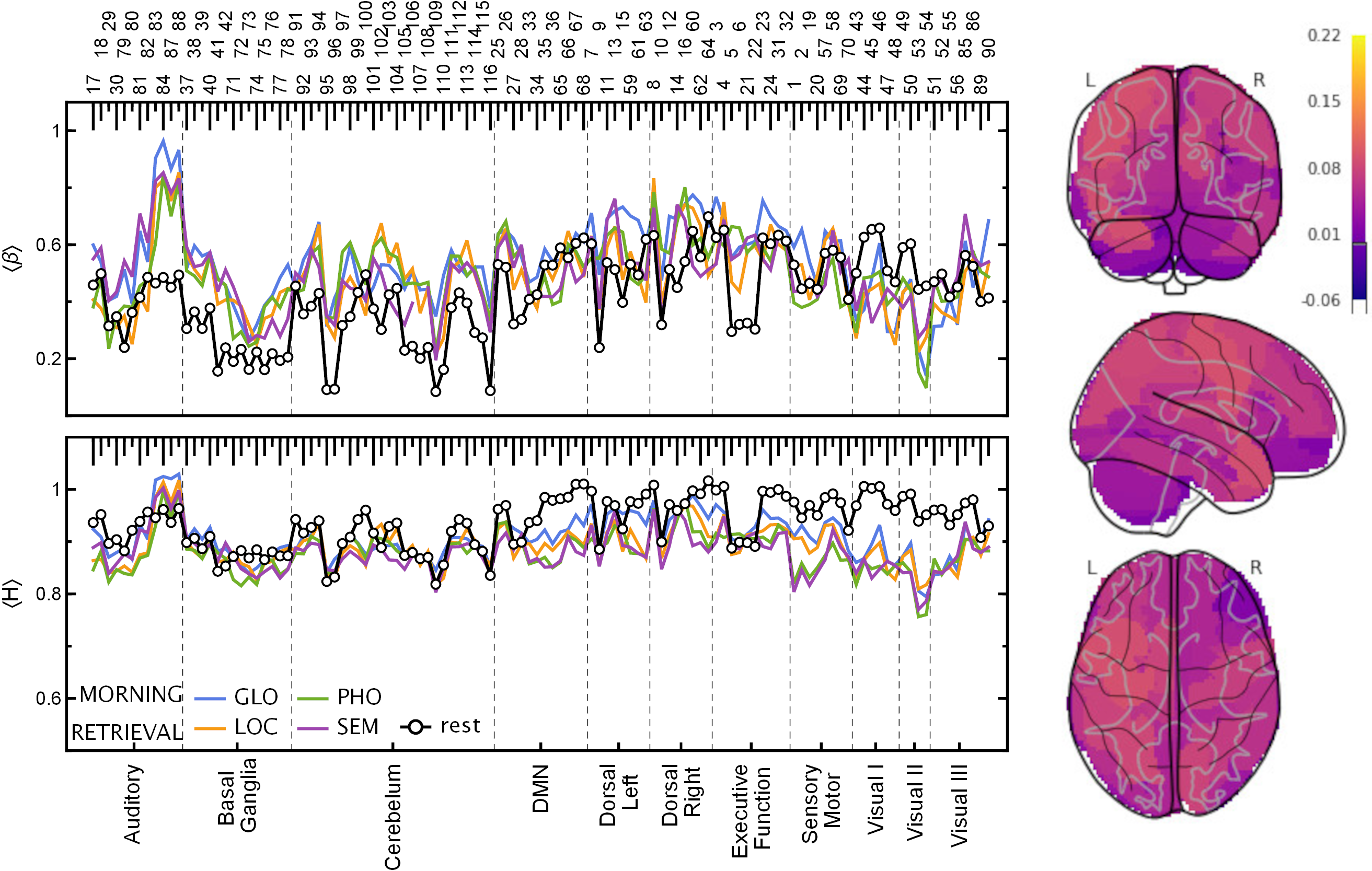}
\caption{(Left panel) Plots of the spectral, $\beta$, and Hurst, $H$, scaling exponents estimated for time series in the retrieval phase. The exponents were calculated for each ROIs and ordered on the plot according to the AAL atlas (top labels) and  resting-state networks (bottom labels). (Right panel) Glass brain plot of $\langle H \rangle$ differences between the GLO and PHO tasks rendered with \textit{nilearn}~\cite{nilearn,nilearn_soft,scikit-learn}. For evening session results, see Supplementary Fig.~A.4. 
Abbreviations: visual-verbal tasks (semantic, SEM, and phonological, PHO), visual-nonverbal tasks (local information processing, LOC, and global information processing, GLO), resting-state (rest).
}
\label{fig6:Hurst_Ret}
\end{figure}

In Figures~\ref{fig5:Hurst_Enc} and~\ref{fig6:Hurst_Ret}, the estimated exponents $\beta$ and $H$ in the encoding and retrieval phases are presented.
There are clear differences from the results obtained for the entire time series.
The $H$ values in the segmented tasks are higher than in the entire ones.
On the other hand, the values of the resting state $H$ obtained from similarly preprocessed data (cf.
Section~\ref{sec:Data_Preprocessing}) reveal a slightly weaker correlation level than for the original data.
However, the most exciting observations come from comparing the scaling properties of different experimental phases.
All tasks generally decreased $H$ exponents with respect to the resting state (by $0.049$ on average, PHO and SEM more than GLO and LOC), and even more so in the encoding phase (by another $0.020$, especially PHO and SEM, then GLO but without interaction for LOC).
For the encoding phase, the Hurst exponent variation is greater than those estimated for the retrieval phase.
The variance of $H$ due to ROIs is more than twice the residual variance of the mixed model,
indicating that finding interactions between tasks and individual ROIs should be the next step.

For the AAL regions belonging to the Visual and Sensorymotor RSNs, the values of $H$ drop significantly.
This effect is especially strong for verbal tasks (SEM and PHO) for which $H \approx 0.5$ (Visual II) and $H \approx 0.7$ (Motor), which is characteristic of uncorrelated and moderately positively correlated data, respectively.
It is not as conspicuous in the case of visuospatial tasks (GLO and LOC) where a significant difference between resting state and task is mainly visible for Visual RSN.
The Hurst exponent's highest values indicate the $1/f$ process identified for the Auditory RSN, similarly to the resting state data.
The $H$ values in Visual RSN are higher in the memory retrieval phase than in encoding; however, there is still a noticeable decrease with respect to the resting state.
Moreover, in retrieval, there is hardly any difference between verbal and visuospatial tasks.
We have found no significant effect of time-of-day (TOD, morning and evening sessions), but there was one significant positive interaction of TOD with the PHO task.
The code for and full output of the statistical model are provided in the Supplementary Listing 1 and Table C.1.

\begin{figure}
\centering
\resizebox{0.95\textwidth}{!}{
\includegraphics{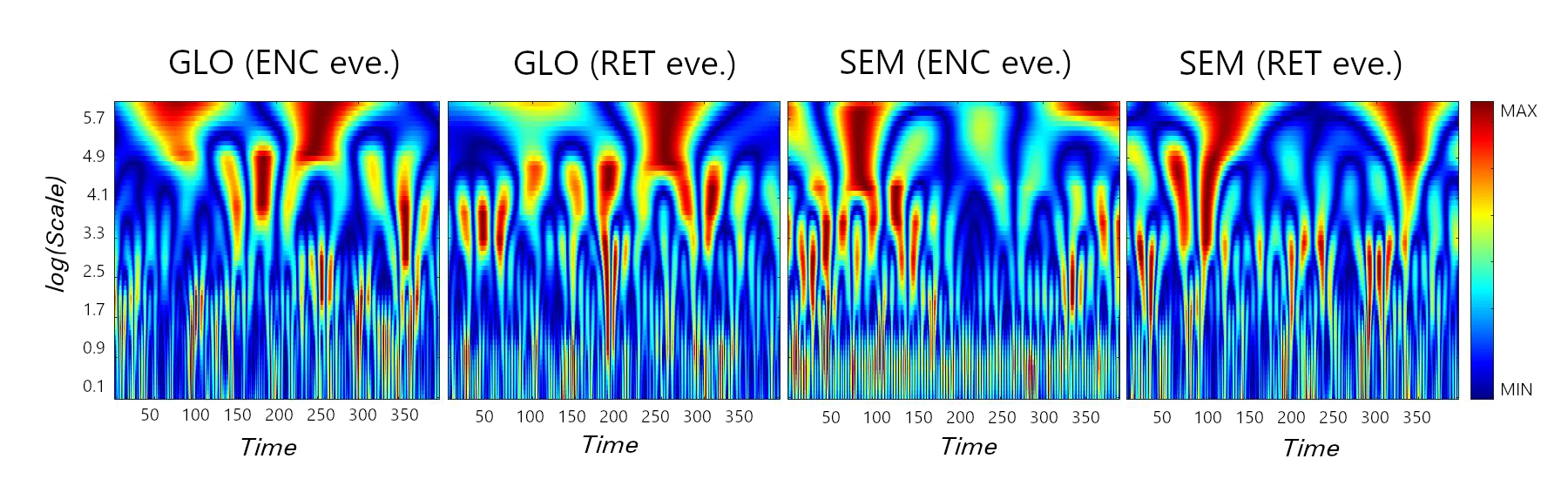}
}\caption{Scalogram of the signals from ROI no. 54 (Visual II RSN) in evening (eve.) for two tasks, GLO and SEM, in memory encoding and retrieval phases. The $W_{\psi}$ amplitude is coded by 256 colours. Abbreviations: experimental phases (encoding, ENC and retrieval, RET), visual-verbal tasks (semantic, SEM), visual-nonverbal tasks (global information processing, GLO).}
\label{fig7:WT}
\end{figure}

To illustrate the difference in the hierarchical structure of the analysed signals, we applied the wavelet transform.
This method is capable of revealing the self-similar properties of the signals, which can be presented as a heat map on a space-scale half-plane called a scalogram.
The scalogram in Fig.~\ref{fig7:WT} represents the sample signals (ROI no. 54, Visual II RSN, TOD evening) for which the most evident differences between the scaling properties of two tasks, GLO and SEM, appear with respect to the experimental phase.
The $W_{\psi}$ amplitude is coded by 256 colours.
It is easy to notice the hierarchical structure underlying the fluctuations.
The wavelet transform produces a kind of branching structure with large regions of high amplitude on a large scale and a much denser distribution of singularities on the smaller one.
Moreover, for the semantic task in the encoding phase, the concentration of the high amplitude on a small scale is also noticeable.
It indicates the high volatility of the signal and the strong singularities quantified by small values of scaling exponents $H$.

\subsection{Cross-correlations study}
\label{sec:res_CC}

\begin{figure}
\centering
 \resizebox{0.5\textwidth}{!}{
\includegraphics{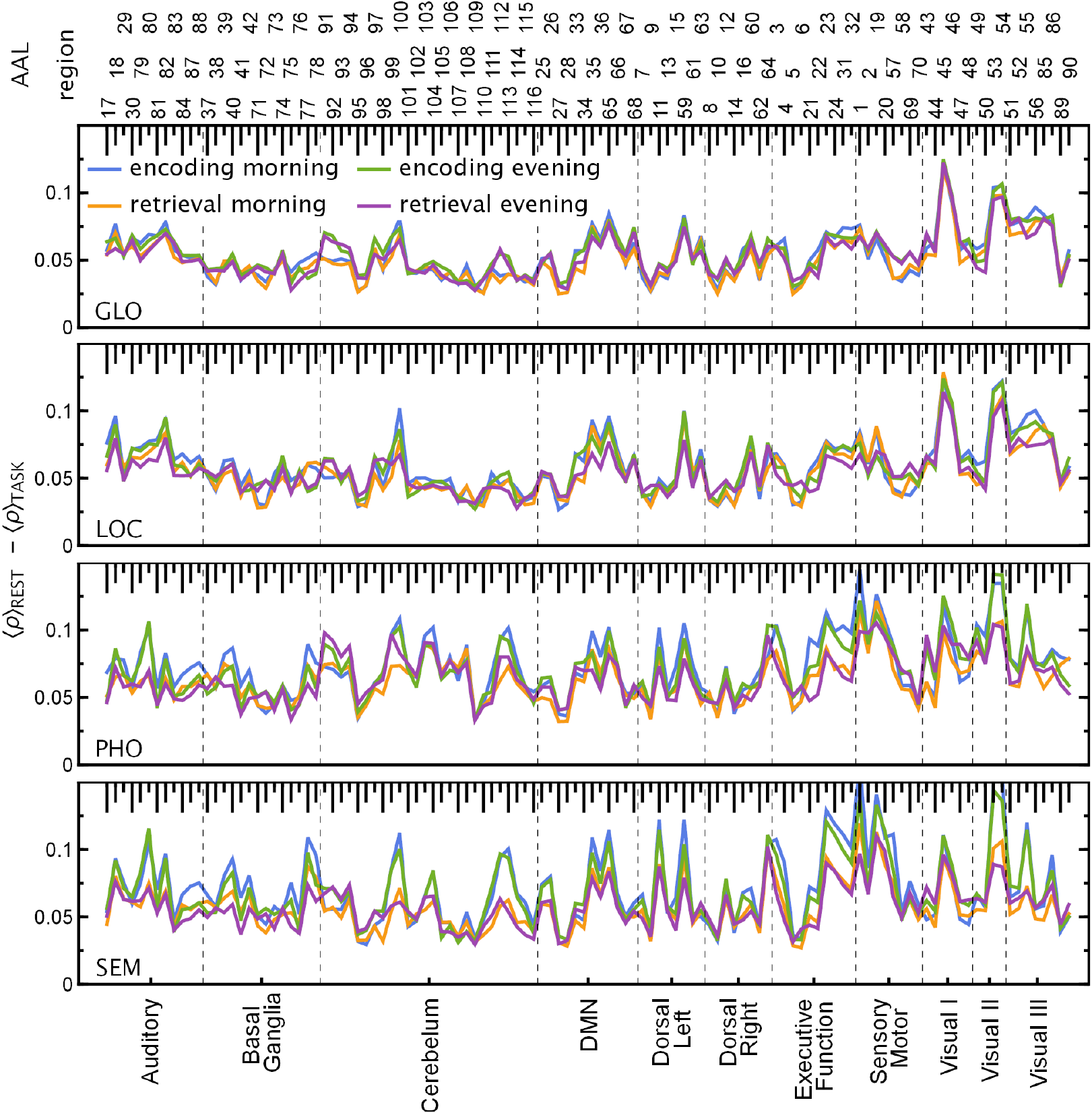}
 }
\caption{Absolute differences between $\langle\rho(q=1,s=10)\rangle$ (group mean for each ROI) for tasks GLO, LOC, PHO, SEM and resting state. Abbreviations: visual-verbal tasks (semantic, SEM, and phonological, PHO), visual-nonverbal tasks (global information processing, GLO, and local information processing, LOC).}
\label{fig2:diffs}
\end{figure}

From the perspective of cross-correlation analysis, we can identify which ROIs are the most correlated with the other ones. 
To measure it, we averaged $\rho(q=1,s=10)$ over all subjects;
next, we calculated the absolute difference of each element of the correlation matrix between the four tasks and resting-state;
finally, for a given ROI we averaged the difference over all the other ROIs, $\langle\rho(q=1,s=10)\rangle$;
the results are presented in Fig.~\ref{fig2:diffs}.

In the case of visuospatial tasks (GLO and LOC), the largest difference from the resting state is reached for Visual I and Visual II RSNs.
For the verbal tasks (PHO and SEM), again the largest difference is for the Visual II region, but also for the sensory-motor and part of the executive areas, especially in the encoding phase of the experiment.

\begin{figure}
\centering
\resizebox{.95\textwidth}{!}{
 \includegraphics{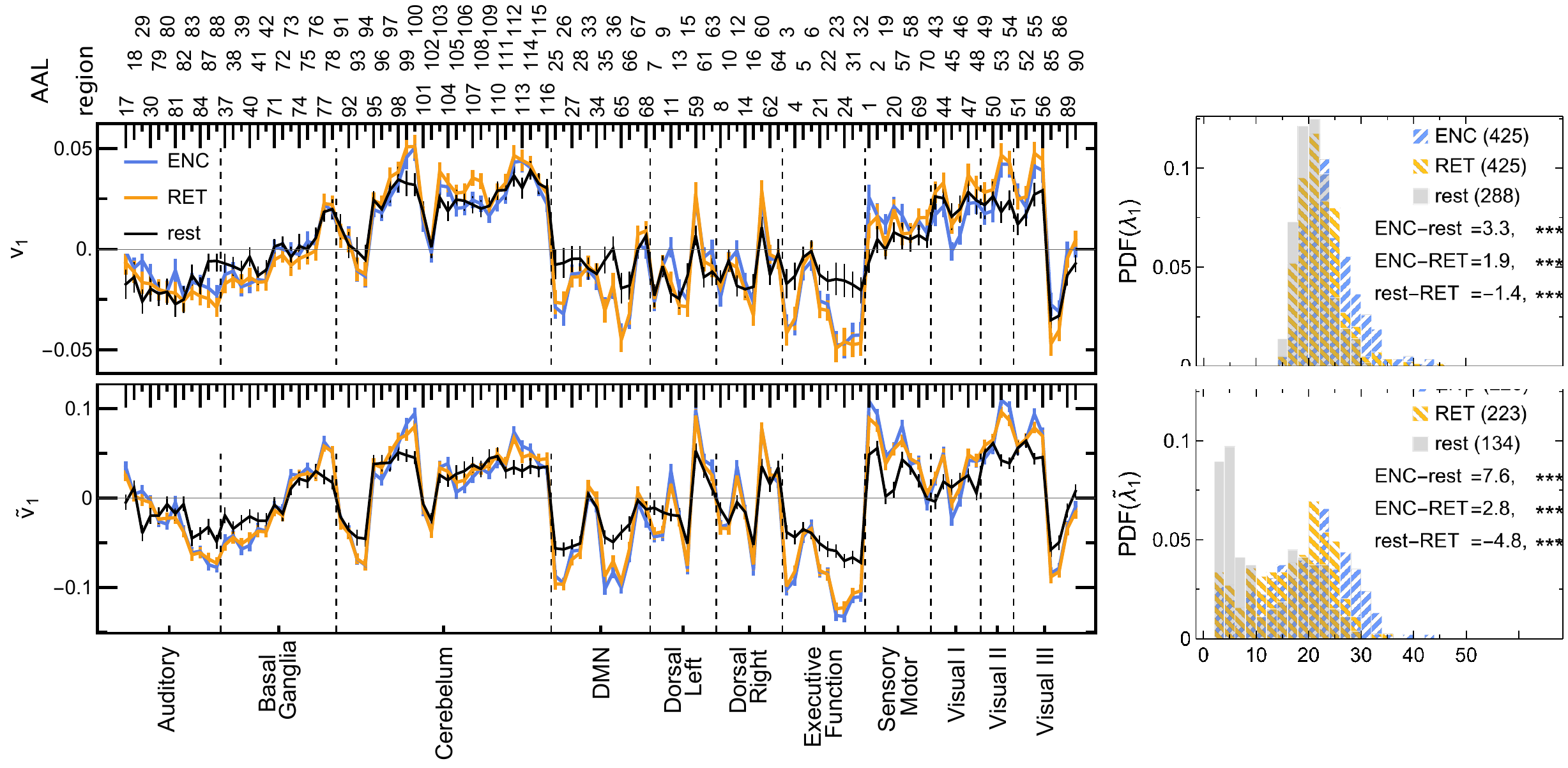}
}
\caption{Average elements of (top) the principal eigenvector and (bottom) the first aligned eigenvector of $\rho$ correlation matrices for encoding, retrieval, and resting state. The bars mark standard errors over all subjects, tasks, and conditions. Histograms of eigenvalues of (left) the principal eigenvector and (right) the first aligned eigenvector of nonlinear correlation matrices for encoding, retrieval and resting state pooled over all subjects, tasks and conditions. The values below the legend indicate the location differences between pairs of distributions. Abbreviations: experimental phases (encoding, ENC and retrieval, RET), resting-state (rest).}
\label{fig4:Pq_eigenvecs1}
\end{figure}

We also checked whether the largest eigenvalues calculated for the correlation matrix of 116 ROIs in the task were significantly different from the resting state.
As expected, naturally ordered eigenvalues, $\lambda_i$, showed much weaker differences, since they were associated with noisier eigenvectors than the clustered eigenvalues, as visible in Fig.~\ref{fig4:Pq_eigenvecs1}.
Henceforth, we focus only on the clustered eigenvalues, $\tilde{\lambda}_i$.
The eigenvalues of Pearson correlations showed fewer differences, which appeared dispersed among several eigenvalues and had smaller magnitudes (for instance, when testing for all pairwise differences between GLO, LOC, PHO, SEM and resting-state,  $\tilde{\lambda}_i$ with $i=5, 11, 13$ showed 2-4 significant differences each and their magnitudes lied between 0.9 and 2.5).
From the perspective of $\rho(q=1,s=10)$ coefficient, taking into account also nonlinear correlations, more differences appeared, they were localised in larger eigenvalues and had larger magnitudes (in the same pairwise comparisons as above, $\tilde{\lambda}_1$ contained six differences of magnitude between 4.0 and 8.8, $\tilde{\lambda}_2$ four differences ranging between 3.4 and 4.6, and many smaller ones for $i=3,4,5,7,9,10,15$).
There were significant differences between morning and evening; however, they appeared only in three eigenvalues for Pearson and two for $\rho(q=1,s=10)$ correlations.
For details, see Supplementary Tables~B.2-7.

In particular, Fig.~\ref{fig4:Pq_eigenvecs1} (bottom) shows that the largest clustered eigenvalue indicates differences not only between the resting state and the tasks, but also between the encoding and retrieval phases.
Similarly, the largest clustered eigenvalue shows differences not only between the resting state and GLO, LOC, PHO, and SEM (4.5***, 4.0**, 6.6*** and 8.8***, respectively), but also between SEM and both GLO and LOC (-4.3**, -4.8***).
Such differences (either the tasks vs resting state or GLO vs PHO and SEM, and LOC vs PHO) repeat also in other eigenvalues, although they are fewer per eigenvalue.

In Fig.~\ref{fig4:Pq_eigenvecs1} we show an example of an eigenvector associated with the largest eigenvalue of the $\rho$ correlation matrix.
Clustering of the eigenvectors was effective, as reflected by the smaller error bars compared to the unclustered principal eigenvectors.
This eigenvector encodes, for instance, opposite activations of such RSNs as Executive Function and DMN versus Sensorymotor and Visual II, with some other RSNs visibly divided, e.g., Visual III or Basal Ganglia.
The principal eigenvector also shows less variance in the resting state than in encoding and retrieval phases.
Particularly more active in tasks are the Executive Function, DMN, Sensorymotor, Visual II and III RSNs.
Even though the pattern between encoding and retrieval is almost identical, the minima and maxima have a systematically greater spread in encoding, reflecting, among others, their greater consistency across subjects and tasks.
Naturally, the eigenvectors corresponding to the other results in Supplementary Tables~B.2-5
encode subtler effects.
They show different patterns of ROI contributions, and each will need a separate interpretation and validation.


\section{Discussion}
\label{sec:con}


Analysis of BOLD signal scaling properties reveals significant differences between task performance and resting state.
Resting-state signals are more regular than those related to task phases (see Fig.~\ref{fig4:Hurst_entire_data}).
This observation is quantitatively described by the Hurst exponents $\langle H \rangle\approx 1.1$ for resting state and $\langle H \rangle\approx 0.7$ for tasks.
The most distinct BOLD signals are related to Auditory (highly correlated), AAL ROI 83-84 (left and right superior temporal poles), and 87-88 (left and right middle temporal poles), and Visual II (minimally correlated), ROI 53-54 (left and right inferior occipital gyrus).
The temporal pole is a heterogeneous structure and many studies have demonstrated its role in various cognitive functions~\cite{herlin_2021}.
It is engaged in both visual memory processing~\cite{roland_1990} and semantic functions~\cite{jouen_2015,jouen_2018}.
Previous research confirmed the dense connections between the dorsolateral temporal pole and the medial temporal cortex and proved the involvement of the temporal pole in memory~\cite{corcoles_2019}.
The inferior occipital gyrus belonging to the Visual II network participates in primary visual processing, as well as word identification as part of the visual word form area~\cite{vigneau_2005}.
These results are also consistent with the study by Churchill et al.~\cite{churchill_2016}, who showed that cognitive effort decreases fractal scaling throughout the brain.
It should be noted that the value of $H$ in resting-state indicates $1/f$ process commonly identified in nature.

However, the most intriguing results come from the analysis of the encoding and retrieval phases.
The Hurst exponents estimated especially for the Visual II network are low compared to the other ROIs, whereby the differences are the most significant for the encoding phase.
Visual cortex activity is associated with the processing of stimuli in working memory tasks and has been shown to distinguish correct and incorrect stimuli at both encoding and retrieval~\cite{pessoa_2002}.
In retrieval, parts of the early visual processing areas had been found to be preferentially linked to correct recognition when compared to false ones, and these differences were linked to a sensory reactivation of memorised stimuli~\cite{slotnick_2004,slotnick_2006}, namely, true recognitions were associated with more sensory details than false ones.
Similarly, visual cortex activity had also been linked to imagery retrieval of sensory details when recognising information from memory~\cite{dennis_2014}.
Taking these findings into consideration, the differences in $H$ values between our tasks (the lowest $H$ in visual-verbal tasks and particularly high differences during encoding) could be linked to a larger amount of sensory details (to be encoded and then retrieved) in visuospatial stimuli than in verbal material.
Based on the results obtained, it is easy to distinguish between the experimental phase (encoding/retrieval) and the task type (visual-verbal/nonverbal).
The recent EEG study~\cite{wolff_2020} provides evidence for dynamics of encoding and maintenance processes of working memory. The results show a temporally stable coding scheme with possible dynamics during working memory maintenance. The results of our study using fMRI and fractal and spectral analysis indicate differences in dynamics within encoding and retrieval phases. Both studies confirm a dynamic-processing model of working memory and demonstrate the necessity of applying the new methodological and computational approaches.

According to the detrended cross-correlation analysis, $\rho(q)$,
the most significant differences in the level of correlation as compared to resting-state
are in the areas of Visual I and Visual II for the visuospatial tasks
and in the Visual II and Sensorymotor networks for the verbal tasks.
The above differences are more visible when correlations are measured with the coefficient $\rho (q,s)$, which, in addition to linear correlations, can identify nonlinear relationships.
From this perspective, the most outstanding networks are DMN (default mode network), Sensorymotor, Cerebellum, and Visual II (ROI 53,54 -- inferior occipital gyrus).
DMN consists of the posterior cingulate cortex, angular gyrus, and precuneus and is traditionally thought to be active while mind-wandering at wakeful rest.
However, more recent studies showed its involvement during thoughts about task performance~\cite{sormaz_2018}.
The precuneus had been found to be involved in the encoding and retrieval processes of episodic memory, executive functions, and visuospatial imaginary (for a review, see:~\cite{cavanna_2006}).
Rao and colleagues (2003)~\cite{rao_2003} showed that both the angular gyrus and the precuneus are activated during the spatial location information processing.
Furthermore, some studies confirmed the involvement of the precuneus in attention~\cite{simon_2002}.
The cerebellum, which is similar in structure to the cerebral cortex, has its own functional architecture, where individual lobules are involved in various motor and cognitive functions (for a review, see:~\cite{stoodley_2009}).
Among others, lobule VI and vermis had been reported to be activated during spatial, attentional, and working memory tasks~\cite{kuper_2015,brissendsen_2019}.
Our results are in line with the studies mentioned above, showing differences in correlations in lobule VI and vermis of the cerebellum.
Sensorimotor network (ROI: 1, 2 -- left and right precentral gyrus, 57, 58 -- left and right postcentral gyrus, and 69, 70 -- left and right paracentral lobules) is activated primarily during processing of sensory and motor information.
However, there is evidence that the precentral gyrus is involved in the retrieval of false memories~\cite{kurkela_2016}.
Both the precentral and postcentral gyri were also previously reported in a similar visuospatial working memory to show correctness-related differences in activity when younger adults processed a lure~\cite{sikora_wachowicz_2021}.
The one limitation of the presented analysis is that it does not allow distinguishing between positive responses and false memories. Another technique based on extracting short significant fMRI events can be used for this purpose \cite{Ceglarek_rbeta_2021}.

\section{Conclusions}
\label{sec:summary}

To our knowledge, this is the first time that spectral and fractal analysis were used to investigate the encoding and retrieval processes of the working memory.
The presented results clearly show that the temporal organisation of the fMRI time series contains distinctive information about brain activity related to different types of tasks and phases of working memory.
Analysis of the scaling properties of the signals revealed that those corresponding to task performance are significantly different from those in the resting state. The differences in signals between encoding and retrieval phases showed that the used methods of analysis are sensitive to dynamic changes in these processes.
The results of the fractal and spectral analysis presented in our study -- consistent with previous research related to visuospatial and verbal information processing, working memory (encoding and retrieval), and executive functions -- shed new light on these processes.

\section{Data} 
\label{sec:Data} 

Below, in Sections~\ref{sec:Data_Participants}-\ref{sec:Data_Tasks} we provide a summary of the data from the experiment described at length in~\cite{Lewandowska2018}. 
Details on fMRI data acquisition and preprocessing can be found in~\cite{fafrowicz_beyond_2019}. 
In Section~\ref{sec:Data_Preprocessing} we provide processing details specific to the current paper.

\subsection{Participants} 

\label{sec:Data_Participants} 

The volunteers first underwent an online sleep-wake assessment that included night sleep quality – Pittsburgh Sleep Quality Index (PSQI)~\cite{buysse_pittsburgh_1989}and daytime sleepiness – Epworth Sleepiness Scale (ESS)~\cite{johns_new_1991}, and completed sleep-wake assessment diurnal preference as measured by the Chronotype Questionnaire~\cite{oginska_chronotype_2017}. 
Exclusion criteria were: age below 19 or above 35, left-handedness, psychiatric or neurological disorders, dependence on drugs, alcohol or nicotine, shift work or travel involving moving between more than two time zones within the past two months, ESS score above 10 points. 
Next, they were qualified for the study with genetic testing for the polymorphism of the clock gene PER3, which was considered a hallmark of extreme diurnal preferences~\cite{archer_length_2003}. 
Finally, 54 subjects were selected (32 females, mean age: 24.17±3.56 y.o.) divided into 26 morning-oriented and 28 evening-oriented types based on the Chronotype Questionnaire and PER3 genotyping. 

The volunteers signed an informed consent form and were remunerated for their participation in the experiment.
The study was carried out in accordance with the Declaration of Helsinki and approved by the Research Ethics Committee at the Institute of Applied Psychology at the Jagiellonian University.

\subsection{Tasks} 

\label{sec:Data_Tasks} 


Data were collected at two times of the day (TOD; morning and evening), four experimental visual tasks (verbal: semantic and phonological, `SEM' and `PHO', and nonverbal: processing of global or local characteristics of abstract objects, `GLO' and `LOC'), two phases of each task (encoding, `ENC', and retrieval, `RET') and resting state (`rest`). 
The four tasks were based on the Deese–Roediger–McDermott paradigm~\cite{deese_prediction_1959,roediger_creating_1995} and specifically the version dedicated to studying short-term/working memory distortions~\cite{atkins_neural_2011}. 
In the semantic task, the stimuli were Polish words matched by semantic similarity. In the phonological task, meaningless pseudowords characterised by phonological similarity were used. On the other hand, in the global processing task, the stimuli were abstract figures requiring holistic processing (differing in the number of overlapping similarities). In the local processing task, the stimuli differed in one specific detail and thus required local processing.

In each task, participants were asked to memorise sets of stimuli (encoding phase, 1.2-1.8 s) which were followed by either simple distractors (in verbal tasks, 2.2 s) or masks (in visuospatial tasks, 1.2 s). Then (after about 6.1 s), another single stimulus, called \textit{probe}, appeared on the screen (2 s).
The probe stimulus came in one of three possible types: \textit{positive} (it had appeared in encoding), \textit{negative} (it was new and dissimilar to the encoding ones) or \textit{lure} (it was new but associated with or similar to one of the encoding stimuli).  
At this point (retrieval phase), participants were asked to answer whether the stimulus had appeared in the preceding set (with the right-hand key for "yes" and the left-hand key for "no" responses). After an 8.4-s intertrial interval, a fixation point (450 ms) and a blank screen (100 ms) were presented, followed by a new set. There were 60 sets of stimuli followed by 25 positive probes, 25 lures, and 10 negative probes in each task. Sample experimental tasks are depicted in Supplementary Fig. A.1. The behavioural data on the response accuracy and reaction times for each task and probe type are presented in Supplementary Table~B.1.

Each participant performed the tasks twice: during morning and evening sessions, two versions of each task were created and presented in a random way, and the order of the presented stimuli was randomised as well.
The resting state with eyes open was recorded for 10 minutes. 
One day before the study was conducted, participants were familiarised with the experimental procedure and underwent a training session.
Further details of the experimental paradigm can be found in~\cite{Lewandowska2018}. 

\subsection{Data preprocessing} 

\label{sec:Data_Preprocessing} 

\begin{figure}
\centering
{
\includegraphics[width = 0.8 \textwidth]{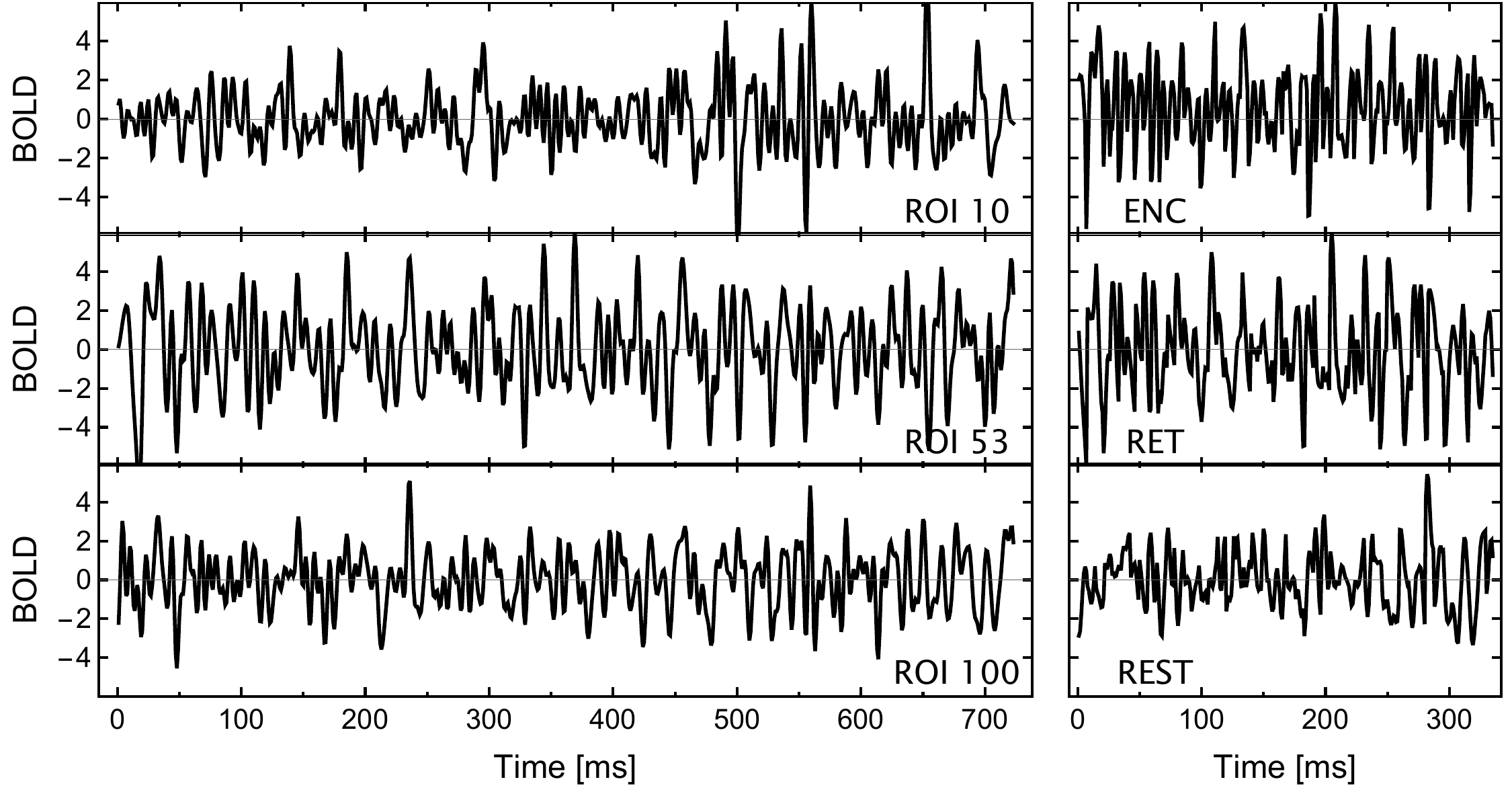}
}
\caption{Examples of BOLD time series (left) from ROIs located in Dorsal Right, Visual II and Cerebellum RSNs during semantic task and (right) from ROI 53, Occipital\_Inf\_L, during memory encoding and retrieval in the same task, and resting state.}
\label{fig2:Entire_data}
\end{figure}

 BOLD signals were collected at repetition time TR $ = 1.8\textrm{ s}$ with a 3T Siemens Skyra MR System and underwent the standard preprocessing described in~\cite{fafrowicz_beyond_2019}. 
The signals were then averaged within 116 brain regions (Regions of Interest, ROIs) from the Automated Anatomical Labeling (AAL1) brain atlas~\cite{tzourio-mazoyer_automated_2002}. 
For a given experimental session, one can form the data in a matrix with 116 columns corresponding to the ROIs and the number of rows corresponding to the length of the time series. Examples of the time series are shown in Fig.~\ref{fig2:Entire_data} (left panel).

To highlight possible differences between the encoding and retrieval phases, we additionally preprocessed the time series as follows.
For each experimental session, from the ROI time series matrix, all data segments related to the encoding (retrieval) phase were extracted and concatenated, in order of appearance, into a new matrix.
Each segment started at the stimulus onset and was $~10 s$ long (6-7 TRs), which approximately corresponds to the time width of the peak of the haemodynamic response function.
In the case of encoding, the segments included presenting the distractor, and in the case of retrieval, they included the intertrial interval. 
For a given session, concatenation of segments from all 60 stimuli resulted in 400-TR long time series. 
To evaluate the statistical significance of the results, we also analysed the time series of resting state with eyes open preprocessed similarly to the task data.
Therefore, in this case, the time series length of 400 time points was cut into nonoverlapping segments of $~10\textrm{ s}$ (the same length as in the task), which were then randomly shuffled, see Fig.~\ref{fig2:Entire_data} (right panel).
Thus, we considered the time series corresponding to the memorisation, retrieval, and resting-state phases separately.  

\section{Methods}
\label{sec:meth}

Analysis of the time series' correlation structure is one of the primary methods for identifying patterns or dependencies in the data and inferring the system's organisation responsible for generating the signal. 
We decided on applying two techniques of autocorrelation time series analysis, namely, detrended fluctuation analysis (DFA)~\cite{peng_mosaic_1994} and power spectral density (PSD)~\cite{welch_1967}. 
These two complementary methods can quantitatively describe the temporal multiscale organisation of the data and, hence, its fractal properties.
Moreover, to qualitatively present the hierarchical structure of the time series, we applied wavelet-based multiresolution analysis. 
To characterise cross-correlations between the time series, matrices of the $q$-dependent detrended correlation coefficients (DCC) were analysed~\cite{kwapien_detrended_2015}. 
The diagram representing the consecutive steps of the data preprocessing and analysis is depicted in Fig.~\ref{fig1:flowchart}.  

\subsection{Fractal analysis of the time series}
\subsubsection{Detrended fluctuation analysis}
\label{sec:meth_H}

One of the most reliable and commonly used methods to detect the self-affinity of the signal is the detrended fluctuation analysis proposed by Peng et al.~\cite{peng_mosaic_1994} to analyse DNA. The algorithm consists of the following steps. 
For a time series $x_i$ of length $N$ ($i=1,2,\ldots,N$), first, calculate the profile
\begin{equation}
X\left(j\right) =\sum_{i=1}^j[x_{i}-\langle x \rangle],
\label{profiles_DFA}
\end{equation}
where $\langle . \rangle$ denotes the average of $x_i$ over the entire time series.
Next, the profile is divided into $2N_s$ nonoverlapping segments of length $s$ (where $N_s=\lfloor N/s \rfloor$ and $\lfloor . \rfloor$ denotes the floor function), numbered by an index $\nu$, starting from the beginning and the end of the time series.
The local trend is estimated in each part $\nu$ by fitting a polynomial $P^{(m)}_{\nu}$ of the order $m$ to the data and then is subtracted.
The polynomial order $m$ is a parameter of the method.
According to the study~\cite{oswiecimka_effect_2013} a reasonable choice is $m=2$, which we used in our analysis.
In each segment, the detrended variance is calculated 
\begin{equation}
F^2(\nu,s)=\frac{1}{s}\sum_{k=1}^{s}\left(X\left((\nu-1)s+k\right)-P^{(m)}_{\nu}(k)\right)^2.	
\label{detrended_variance}
\end{equation}
Finally, the scale-dependent fluctuation function is estimated 
\begin{equation}
F(s)=\sqrt{\frac{1}{2N_s} \sum_{\nu=1}^{2N_s}F^{2}(\nu,s)}. 
\label{Fs}
\end{equation}
For a fractal time series, one observes the power-law behaviour
\begin{equation}
F(s)\sim s^H,
\end{equation}
where $H$ denotes the Hurst exponent. Therefore, if the time series is only short-range correlated, the Hurst exponent assumes the value $\sim$ 0.5. In the case of long-range fractal-correlated time series, a deviation of $H$ from 0.5 is observed, respectively,
\begin{itemize}
    \item $0.5<H<1$ for positively autocorrelated data (persistent signal),
    \item $0<H<0.5$ for negatively autocorrelated one (antipersistent signal).
\end{itemize}

\begin{figure}[ptb]
\centering
\includegraphics[width = \textwidth]{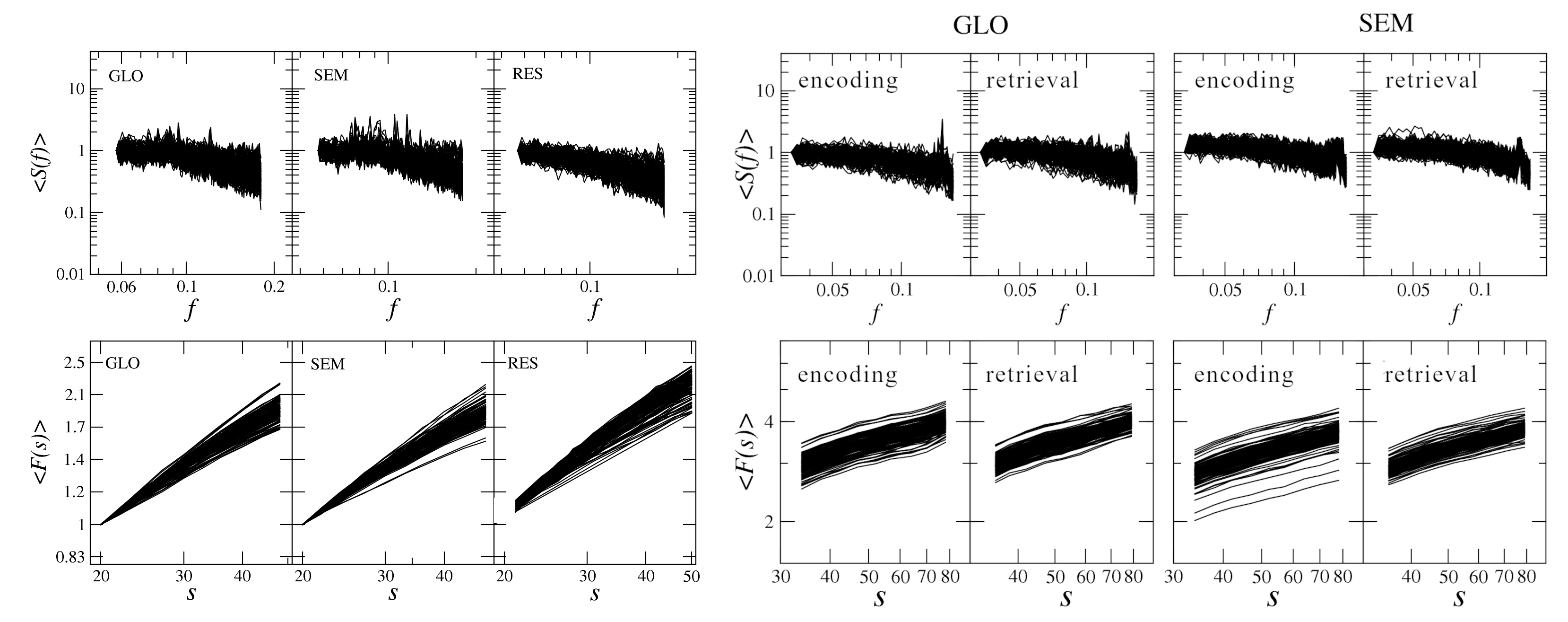}

\caption{The average of PSD, $\langle S(f) \rangle$ and fluctuation functions, $\langle F(s)\rangle$, estimated for two tasks (verbal semantic, SEM, and nonverbal global information processing, GLO) and resting-state (RES). (Left) Computed on entire time series. (Right) Computed after dividing time series into the experimental phases: encoding and retrieval. Each line corresponds to a different ROI.}
\label{fig3:F_entire_data}
\end{figure}

The results for the GLO and SEM tasks (visuospatial and verbal, respectively) and resting-state are presented in Fig.~\ref{fig3:F_entire_data}.
Since the present study focuses on short-term dependencies in the signals, the plots are restricted to a range of small scales.
The presented characteristics clearly follow power laws within the scale range, $s$, considered.
Furthermore, the scaling properties depend on the ROI, which can be easily seen by their different slopes of $\langle F(s)\rangle$ and $\langle S(f) \rangle$ in Fig.~\ref{fig3:F_entire_data} (left panels).
These quantities were recomputed after dividing the tasks into the experimental phase (encoding and retrieval) averaged over all participants, as shown in Fig.~\ref{fig3:F_entire_data} (right panels). 
Even though the scaling length after the additional processing is shorter than the results for the full-length data, the scaling proprieties can still be identified.

\subsubsection{Power spectrum and wavelet transform analysis}

One of the main mathematical tools used to identify quasiperiodicity in time series is power spectrum analysis.
It is able to characterise the linear correlations of the time series via the distribution of its variance (energy) at specific frequencies.
The spectrum of the time series $x_j$ of length $N$ ($j=0,1,\ldots,N-1$) is estimated via the square of the Fourier transform modulus: 
\begin{equation}
S(f)=\left|\sum_{j=0}^{N-1}x_je^{-2\pi ifj/N}\right|^2, 
\end{equation}
where $f=0,1,\ldots,N-1$.
In the case of fractal signals, the scaling relation is observed
\begin{equation}
S(f)\thicksim 1/f^\beta,
\end{equation}
where $\beta$ is a spectral scaling exponent related to the Hurst exponent via the relation
\begin{equation}
\beta = 2H-1.
\label{eq:relation_H_beta}
\end{equation}
Thus, DFA and PSD can be considered complementary methods of quantifying linear long-range correlations in fractal time series.   

To emphasise differences in the hierarchical organisation of the time series analysed, we applied wavelet analysis~\cite{arneodo_1995}.
By its virtue, the time series is decomposed into different periodic components taking into account the time of their occurrence.
Thus, the wavelet transform can be considered a mathematical microscope that provides insight into the intrinsic organisation of the time series.
The wavelet transform for discrete signals is defined as
\begin{equation}
W_{\psi}\left[s,k\right]=1/\sqrt{s}\sum_{j=1}^{N}x_j\psi\left(j-k/s\right),
\end{equation}
where $\psi$ is the wavelet well localised in the time and frequency domain, $s$ denotes the scale parameter, and $k$ is the wavelet's position in time. 
In this work, we used the Mexican hat (the second derivative of the Gaussian) as a wavelet. 

\subsection{Cross-correlation analysis}
\label{sec:meth_CC}
To describe the nonlinear cross-correlations, in addition to the standard Pearson correlation coefficient~\cite{pearson_note_1895}, we used $q$-dependent DCC, $\rho(q,s)$~\cite{kwapien_detrended_2015}.
It allows one to quantify the cross-correlations between two time series $x_i$ and $y_i$ both at particular time scales $s$ and with respect to the amplitude of fluctuations filtered by $q$.  
The procedure of estimating $\rho(q,s)$ extends the formulas (\ref{detrended_variance}) and (\ref{Fs}) to the case of the covariance function.
Given the profiles $X\left(j\right)$ and $Y\left(j\right)$ of the time series [cf. (\ref{profiles_DFA})] the covariance is estimated as
\begin{equation}
\begin{split}
F_{xy}^{2}(\nu,s)=\frac{1}{s}\sum_{k=1}^{s} 
\left(X((\nu-1)s+k)-P^{(m)}_{X,\nu}(k)\right)  
\times 
\left(Y((\nu-1)s+k)-P^{(m)}_{Y,\nu}(k)\right),
\label{Fxy2}
\end{split}
\end{equation}
for segments $\nu=1,\ldots,N_s$ and   
\begin{equation}
\begin{split}
F_{xy}^{2}(\nu,s)=\frac{1}{s}\sum_{k=1}^{s} \left(X(T-(\nu-N_s)s+k)-P^{(m)}_{X,\nu}(k)\right)
\times 
\left(Y(T-(\nu-N_s)s+k)-P^{(m)}_{Y,\nu}(k)\right),
\label{Fxy2T}
\end{split}
\end{equation}
for $\nu=N_s+1,\ldots,2N_s$, where $P^{(m)}_{X,\nu}$ and $P^{(m)}_{Y,\nu}$ denote polynomials fitted to the $X(j)$ and $Y(j)$ profiles, respectively.
Then, $q$-th order covariance function~\cite{oswiecimka_detrended_2014} is
\begin{equation}
F_{xy}^{q}(s)=\frac{1}{2N_s}\sum_{\nu=1}^{2N_s} {\rm
sign}\left(F_{xy}^{2}(\nu,s)\right)\left|F_{xy}^{2}(\nu,s)\right|^{q/2},
\label{Fq}
\end{equation}
where ${\rm sign}(F_{xy}^{2}(\nu,s))$ denotes the sign of $F_{xy}^{2}(\nu,s)$.
The parameter $q$ plays the role of a signal amplitude filter.
Finally, the $q$-dependent DCC, $\rho(q,s)$, is estimated as follows
\begin{equation}
\rho(q,s) = {F_{xy}^q(s) \over \sqrt{ F_{xx}^q(s) F_{yy}^q(s)}},
\label{rhoq}
\end{equation}
where $F_{xx}^{q}$, $F_{yy}^{q}$ are calculated by means of (\ref{Fxy2}) and (\ref{Fxy2T}) when only $X(j)$ or $Y(j)$ is separately considered.
Furthermore, in our calculations we used $q=1$, which ensures that the coefficient $\rho(q,s)$ assumes values within the range $[-1,1]$, making its interpretation analogous to the Pearson correlation coefficient~\cite{kwapien_detrended_2015}. 

\subsection{Eigenvector alignment}
\label{sec:meth_EV}

Having computed the correlation matrix and its eigendecomposition for each configuration of the independent variables $(subject, TOD, task, phase)$,
one would like to statistically compare the distributions of eigenvalues between these configurations.
Na{\"i}vely, the groups of all the largest eigenvalues could be compared,
then the groups of all the second largest eigenvalues, and so on.
However, the order of eigenvalues corresponding to a given correlation pattern may be contingent on interindividual variability.
If the effect of, say, TOD were a shift in the distribution location of a given eigenvalue, then we might as well expect this shift to cause a change in the order of neighbouring eigenvalues.
In such a case, the na{\"i}ve approach would compare the eigenvalue distributions of different correlation patterns.

For these reasons, the eigenvalues were first grouped into sets corresponding to the same correlation pattern.
The grouping was obtained by agglomerative hierarchical clustering of the eigenvectors.
We used Ward's method with squared Euclidean distance $d(u,v)$ modified to allow reflection $\tilde{d}(u,v)=\min(d(u,v),d(-u,v))$, since the eigenvector opposites remain eigenvectors.
The clustering was performed on eigenvectors coming from all possible $(subject, TOD, task, phase)$ configurations associated with large eigenvalues $\lambda > 2$.
This arbitrary cut-off lies above the standard upper fence for outliers ($Q_3+1.5 IQR =1.5$ on average) and results in $13.5\pm1.1$ valid eigenvalues per configuration.
Consequently, the clustering was set to produce 15 clusters.
Additionally, when in a given cluster more than one eigenvector belonging to a given subject had the same $(TOD, task, phase)$ configuration,
only the one associated with the largest eigenvalue was kept, and the others were deleted from further analysis ($24\%$ in total).
The clusters were then ordered according to the average of member eigenvalues.

\subsection{Statistical tests}
\label{sec:meth_ST}
To assess the statistical reliability of the results, we performed an analysis of the randomly shuffled data. This procedure destroys all temporal correlations in the time series but preserves its statistical distribution.
The scaling properties of the surrogates can be quantified by the Hurst exponent $H = 0.5$. 
We also tested a mixed linear model for the global effects of TOD, task, task phase and their interactions, with ROIs as random effects. 
We performed a stepwise model selection by dropping single terms based on $\chi^2$ tests.

We report tests that compare the distributions of correlation matrix eigenvalues in natural order and clustered eigenvalues. Individual data points corresponded to single subjects in a given condition. We used linear models with eigenvalues as the dependent variable and the independent variables: TOD (morning and evening), task (GLO, LOC, PHO, SEM, and resting-state), and task phase (encoding and retrieval; for resting-state we used three independent random samples) and all pairwise interactions. Next, we performed a stepwise model selection by dropping single terms based on F-tests. To assess the differences between levels within the remaining factors, due to the exploratory nature of the study, \textit{post-hoc} we utilised the more conservative Scheffé method~\cite{scheffe1999analysis}. These results are provided below the legends in Fig.~\ref{fig4:Pq_eigenvecs1}, right panels. We use the following significance codes: p<0.001 ‘***’, 0.01 ‘**’, 0.05 ‘*’.
The R code and the data needed to reproduce all statistical tests' results are provided in the Supplementary Materials.

\section*{Acknowledgements }
We would like to thank Prof. Patricia Reuter-Lorenz for her constructive suggestions during the planning and development of the Harmonia project and her valuable support. We also thank Anna Beres and Monika Ostrogorska for their assistance in data collection, Piotr Faba for his technical support on the project and help in data acquisition, and Aleksandra Zyrkowska and Halszka Oginska for help in the process of participant selection. The authors thank Ignacio Cifre for providing the AAL to RSN mapping.

\section*{Author contributions statement}
 A.C., M.F., T.M, K.L, B. S-W conceived and conducted the experiment and aided in data collection. J.O., M.W. and P.O. managed the statistical analysis, analysed the results, wrote and edited the manuscript, and prepared the figures. All authors reviewed the manuscript.

\section*{Funding}
The fMRI data analysis was funded by the Foundation for Polish Science co-financed by the European Union under the European Regional Development Fund in the POIR.04.04.00-00-14DE/18-00 project carried out within the Team-Net programme. The experimental study was funded by the Polish National Science Centre through grant Harmonia (2013/08/M/HS6/00042). 

\section*{Additional information}
\textbf{Competing interests} The authors declare that they have no conflict of interest. 

\section*{Data availability}
The functional MRI data analysed are available from Magdalena Fafrowicz (email: magda.fafrowicz@uj.edu.pl) on reasonable request.


\end{document}